\documentclass[twocolappendix, numberedappendix]{emulateapj}

\shorttitle{How to switch on and off a GRB}
\shortauthors{Bernardini et al.}

\begin{document}

\title{How to switch on and off a Gamma-ray burst through a magnetar}

\author{M.G. Bernardini\altaffilmark{1}, S. Campana\altaffilmark{1}, G. Ghisellini\altaffilmark{1}, P. D'Avanzo\altaffilmark{1}, D. Burlon\altaffilmark{2,3}, S. Covino\altaffilmark{1}, G. Ghirlanda\altaffilmark{1}, A. Melandri\altaffilmark{1}, R. Salvaterra\altaffilmark{4}, S.D. Vergani\altaffilmark{5}, V. D'Elia\altaffilmark{6}, D. Fugazza\altaffilmark{1}, B. Sbarufatti\altaffilmark{1}, G. Tagliaferri\altaffilmark{1}}

\affil{$^1$ INAF -- Osservatorio Astronomico di Brera, via E. Bianchi 46, I-23807 Merate (LC), Italy}
\affil{$^2$ Sydney Institute for Astronomy, School of Physics, The University of Sydney, NSW 2006, Australia}
\affil{$^3$ ARC Centre of Excellence for All-sky Astrophysics (CAASTRO), The University of Sydney, NSW 2006, Australia}
\affil{$^4$ INAF -- IASF Milano, via E. Bassini 15, I-20133 Milano, Italy}
\affil{$^5$ GEPI -- Observatoire de Paris, CNRS UMR 8111, Univ. Paris-Diderot, 5 Place Jules Jannsen, F-92190 Meudon, France}
\affil{$^6$ INAF -- Osservatorio Astronomico di Roma, via Frascati 33, I-00040 Monteporzio Catone (RM), Italy}

\begin{abstract}
One of the most elusive features of Gamma Ray Bursts (GRBs) is the sporadic emission prior to the main prompt event observed in at least $\sim 15\%$ of cases. These precursors have spectral and temporal properties similar to the main prompt emission, and smaller, but comparable, energetics. They are separated from the main event by a quiescent time that may be extremely long and, in some cases, more than one precursor has been observed in the same burst. Precursors are still a puzzle: despite many  attempts none of the proposed models can account for all the observed features. Based on the complete sample of bright long GRBs observed by \textit{Swift} (BAT6), we propose a new scenario for which precursors are explained by assuming that the central GRB engine is a newly born magnetar. In this model the precursor and the prompt emission arise from accretion of matter onto the surface of the magnetar. The accretion process can be halted by the centrifugal drag exerted by the rotating magnetosphere onto the in--falling matter, allowing for multiple precursors and very long quiescent times. 
\end{abstract}


\section{Introduction}

One of the most elusive features of Gamma Ray Bursts (GRBs) is the sporadic emission prior to the main prompt event observed in at least $\sim 15\%$ of cases. Previous studies of precursor activity on BATSE \citep{1995ApJ...452..145K,2005MNRAS.357..722L,2009A&A...505..569B} and \textit{Swift}/Burst Alert Telescope (BAT, \citealt{2005SSRv..120..143B}) samples \citep[][hereafter B08]{2008ApJ...685L..19B} showed that precursors have spectral and temporal properties similar to the main prompt emission, and smaller, but comparable, energetics. They are separated from the main event by a quiescent time that may be extremely long (up to $\sim 100$ s, rest frame), especially if measured in terms of the typical variability time--scale of the prompt emission ($\sim 1$ ms). In some cases, more than one precursor has been observed in the same burst, separated by several tens of seconds.

Current models fail to explain these characteristics. They usually involve the collapse of the progenitor star, either in two--steps, first to a neutron star (the precursor) and then to a black hole (the main emission) \citep{2007ApJ...670.1247W,2009MNRAS.397.1695L}, or via a shock breakout through the stellar envelope preceding the main GRB \citep{2002MNRAS.331..197R,2007ApJ...670.1247W}, or when the fireball becomes transparent \citep{2001ApJ...555L.113R,2000ApJ...530..292M,2007MNRAS.380..621L,2002MNRAS.336.1271D}. These models predict, however, one precursor episode, and in the last two cases, the precursor spectrum should be thermal (i.e. a blackbody), contrary to what is observed. Alternatively, the precursor emission may be related directly to the central engine activity, although this requires ad-hoc conditions \citep{2001MNRAS.324.1147R}.

In this paper we propose a new scenario for which precursors are explained by assuming that the central GRB engine is a newly born magnetar \citep{1992Natur.357..472U,1992ApJ...392L...9D,2007ApJ...669..546U,2011MNRAS.413.2031M}, i.e. a neutron star endowed with a large magnetic field ($B\sim 10^{15-16}$ G). In this model the precursor and the prompt emission arise from the accretion of matter onto the surface of the magnetar. The accretion process can be halted by the centrifugal drag exerted by the rotating magnetosphere onto the in--falling matter, allowing for multiple precursors and very long quiescent times. After the end of the prompt emission, the GRB afterglow can also be influenced by the magnetar, being re--energized by its spin--down power \citep{2011A&A...526A.121D}. 

We made use of the complete sub--sample of bright long GRBs observed by BAT presented in \citet{2012ApJ...749...68S} (hereafter BAT6) to constrain the observational properties of GRBs with precursors in an unbiased way, and to have a direct and independent estimate of  the magnetic field and spin period of the magnetar needed to calculate the characteristic luminosities of the quiescent phase for different GRBs. Among different complete samples available in literature, as the sample of GRBs with optical afterglow promptly observed by GROND \citep{2011A&A...526A..30G} or The Optically Unbiased Gamma-Ray Burst Host (TOUGH) Survey \citep{2012ApJ...756..187H}, the BAT6 sample is the best candidate for the search for precursors. In fact one of its main selection criteria is the brightness of the prompt emission as observed by \textit{Swift}/BAT, providing suitable conditions for the identification and analysis of precursor activity prior to the main event. 

In Section~\ref{sect_bat} we describe the sample selection criteria and the main observational properties of GRBs with precursors in the BAT6 sample. In Section~\ref{sect_propeller} we discuss the model and its main implications. In Section~\ref{sect_rate} we compare the rate of magnetars with the rate of Supernovae Ibc and low--luminosity and normal GRBs. In Section~\ref{conclusions} we draw our main conclusions. We adopt standard values of the cosmological parameters: $H_\circ=70$ km s$^{-1}$ Mpc$^{-1}$, $\Omega_M=0.27$, and $\Omega_{\Lambda}=0.73$. Errors are given at $1\, \sigma$ confidence level unless otherwise stated.

\section{Precursors in the BAT6 sample}\label{sect_bat}

We searched for precursors in all the long GRBs of the BAT6 sample (consisting of $58$ GRBs) adopting the same criteria of B08, where the sample of all the GRBs observed by \textit{Swift} and with redshift measurement was considered, i.e. a precursor is identified if:
\begin{itemize}
\item the peak count-rate of the precursor in the BAT $15-150$ keV energy band light curve is lower than or equal to the main event;
\item the \textit{Swift}/BAT flux falls below the background level before the main event starts.
\end{itemize}
We applied these selection criteria to the \textit{Swift}/BAT light curves binned with a signal-to-noise ratio $S/N=5$, and we identified $10$ GRBs out of $58$ of the BAT6 complete sample ($17\%$; see Table~\ref{tab_bat}) fulfilling these criteria. Six of them are also in the B08 sample. GRB 061222A and GRB 070306 have two clear and well-separated precursors, while GRB 060904A and GRB 080602 have a long-lasting precursor emission with several peaks, although not well-separated. Since we are dealing with a rather small sample, when possible we supported our results with those found in previous works on larger samples.

The \textit{Swift}/BAT data were retrieved from the public archive\footnote{http://heasarc.gsfc.nasa.gov/cgi-bin/W3Browse/swift.pl} and processed with the standard \textit{Swift} analysis software included in the NASA's HEASARC software (HEASOFT, ver. 6.12) and the relevant calibration files. The analysis has been performed in the $15-150$ keV energy band, unless otherwise stated. We extracted background-subtracted light curves binned with $S/N=5$ for the identification of the precursors and main event, and binned at $4$ ms for the identification of the interesting time intervals with \texttt{battblocks}. The precursor and main event spectral analysis was performed with XSPEC (ver. 12.6.1). The spectra were fitted with either a single (PL) or a cutoff power-law (CPL) model. A model with a larger number of free parameters was not required to improve the best fit of the spectra, also due to the limited energy band of the BAT instrument. The results of our analysis are detailed in Table~\ref{tab_bat}.

We find that precursors show very similar temporal behavior to their main event, albeit with a shorter duration, from simple FRED (Fast Rise and Exponential Decay) shapes to complex multi-peaked structures. The peak flux of the precursor is lower than the main event (by definition), with a median value for their ratio $\sim 23\%$, but spanning a wide range: from $6\%$ for  GRB 061121 to $68\%$ for GRB 080602. Similarly, the median value of the ratio between the precursor and main event fluence is $~20 \%$, with the record of GRB 080602 where the precursor fluence slightly exceeds the main event total fluence. Precursors and main events thus have comparable brightness and energetics (see Table~\ref{tab_bat}).

Precursors and main events have also consistent spectral indices (a Kolmogorov--Smirnov test gives a probability $P=0.97$ that they are drawn from the same population, see Table~\ref{tab_bat}), similar to what was found BY B08 with a larger sample of GRBs observed by BAT. The spectra in the BAT energy range are non-thermal, being well fit in all cases by a PL or CPL model. We therefore confirm B08 findings, that a pure thermal spectrum can be excluded in the majority of cases (see also \citealt{2005MNRAS.357..722L} with a large sample of BATSE GRBs with precursors).

\begin{deluxetable*}{cc|ccccc|ccccc|c}
\tablecaption{Summary of the prompt emission properties of GRBs with precursors in the BAT6 sample as observed by BAT ($15-150$ keV). \label{tab_bat}}
\tablewidth{0pt}
\tablehead{\colhead{Name} & \colhead{$z$} & \colhead{$F^{\rm p}$} & \colhead{$f^{\rm p}$} & \colhead{$T_{90}^{\rm p}$} & \colhead{$ t_{\rm pk}^{\rm p}$} & \colhead{$\Gamma^{\rm p}$} & \colhead{$F^{\rm me}$} & \colhead{$f^{\rm me}$} & \colhead{$T_{90}^{\rm me}$} & \colhead{$ t_{\rm pk}^{\rm me}$} & \colhead{$\Gamma^{\rm me}$} & \colhead{$f_{\rm min}$}}
\startdata
050318 & $1.44$ & $2.5$ & $12.6\pm1.5$ & $6.65$ & $0.82$ & $2.11\pm0.24$ & $10.7$ & $22.9\pm1.6$ & $-$\tablenotemark{a} & $29.13$ & $1.90\pm0.08$ & $15.9$\\
050401 & $2.90$ & $74.1$ & $62.0\pm18.0$ & $9.14$ & $0.49$ & $1.51\pm0.09$ & $37.8$ & $99.9\pm7.8$ & $8.00$ & $24.00$ & $1.38\pm0.08$ & $6.6$\\
060210 & $3.91$ & $4.6$ & $7.7\pm2.8$ & $9.65$ & $-224.00$ & $1.55\pm0.26$ & $63.9$ & $23.4\pm2.2$ & $167.16$ & $0.44$ & $1.46\pm0.06$ & $1.6$\\ 
060904A-1 & $-$ & $12.9$ & $5.9\pm1.1$ & $35.90$ & $5.08$ & $1.46\pm0.08$ & $57.4$ & $43.6\pm2.0$ & $37.90$ & $55.80$ & $1.52\pm0.03$ & $-$\\
060904A-2 & $-$ & $7.9$ & $7.9\pm1.1$ & $18.50$ & $30.60$ & $1.64\pm0.08$ & $-$ & $-$ & $-$ & $-$ & $-$ & $-$\\
061007 & $1.26$ & $58.7$ & $56.2\pm5.2$ & $12.79$ & $6.82$ & $1.12\pm0.05$ & $380.7$ & $154.9\pm3.6$ & $54.40$ & $45.19$ & $1.04\pm0.01$ & $5.8$\\
061121 & $1.31$ & $6.6$ & $12.5\pm1.3$ & $7.83$ & $2.04$ & $1.63\pm0.08$ & $133.1$ & $199.5\pm3.7$ & $29.45$ & $74.48$ & $1.31\pm0.03$\tablenotemark{b} & $5.9$\\
061222A-1 & $2.09$ & $2.8$ & $6.1\pm1.4$ & $14.3$ & $1.99$ & $1.41\pm0.14$ & $67.1$ & $79.4\pm1.8$ & $47.20$ & $87.00$ & $1.22\pm0.02$ & $1.3$\\
061222A-2 & $2.09$ & $6.8$ & $18.6\pm1.7$ & $20.00$ & $26.30$ & $1.76\pm0.10$ & $-$ & $-$ & $-$ & $-$ & $-$ & $-$\\
070306-1 & $1.50$ & $3.1$ & $6.5\pm2.0$ & $7.70$ & $-115.60$ & $1.29\pm0.28$ & $38.7$ & $29.5\pm1.4$ & $67.60$ & $98.27$ & $1.64\pm0.03$ & $3.3$\\
070306-2 & $1.50$ & $10.2$ & $6.9\pm2.0$ & $26.00$ & $2.03$ & $1.54\pm0.20$ & $-$ & $-$ & $-$ & $-$ & $-$ & $-$\\
080602 & $\sim 1.4$\tablenotemark{c} & $15.2$ & $19.1\pm3.6$ & $17.66$ & $0.73$ & $1.34\pm0.14$ & $13.1$ & $28.2\pm2.0$ & $9.96$ & $60.10$ & $1.06\pm0.07$ & $-$\\
091208B & $1.06$ & $7.9$ & $37.1\pm6.0$ & $3.95$ & $0.48$ & $1.97\pm0.20$ & $22.9$ & $104.7\pm7.2$ & $5.78$ & $8.79$ & $1.75\pm0.12	$ & $40.2$\\
\enddata
\tablecomments{Name and redshift ($z$), fluence ($F$), $1-$s peak flux ($f$), duration ($T_{90}$), peak time ($t_{\rm pk}$), photon index ($\Gamma$) of the spectrum assuming a power-law model of the precursor(s) (superscript ``p'') and main event (superscript ``me''); minimum accretion flux necessary to penetrate the centrifugal barrier ($f_{\rm min}$; see Appendix~\ref{app_magnetar}). Fluences are given in units $10^{-7}$ erg cm$^{-2}$, fluxes in $10^{-8}$ erg cm$^{-2}$ s$^{-1}$, times in s.}
\tablenotetext{a}{The event data of part of the main emission are not available.}
\tablenotetext{b}{Spectrum fitted with a cutoff power-law model. Cutoff energy is $E_{cpl}=499\pm66$ keV.}
\tablenotetext{c}{Photometric redshift on the bases of the most probable host galaxy association in the XRT error circle \citep{2012A&A...545A..77R}.}
\end{deluxetable*}

\subsection{Are precursors correlated to the main event?} 

The GRBs with precursors and with redshift measurement in the BAT6 sample ($8$ GRBs out of $10$ have spectroscopic redshift\footnote{GRB 080602 has a photometric redshift on the bases of the most probable host galaxy association in the XRT error circle \citep{2012A&A...545A..77R}.}, comprising $10$ precursors) allows us to investigate these properties in the cosmological rest frame. We therefore calculated energies and luminosities in a common rest frame energy band ($E_{\rm min}=15 \,(1+z_{\rm max})$ keV $= 73.6$ keV and $E_{\rm max}=150\, (1+z_{\rm min})$ keV $= 309$ keV). To avoid the inevitable correlation between luminosity (energy) and redshift in a flux--limited sample as the BAT6, we make use of the partial correlation coefficient (\citealt{1979ats..book.....K}; see Eq.~2 in \citealt{2012MNRAS.425..506D}).

We identify a moderate positive correlation between the peak luminosity in the common rest frame energy band of each precursor and the subsequent event, with a partial correlation coefficient $r=0.68$ (null-hypothesis probability $P=0.01$).  A moderate correlation is also found between precursor and the subsequent event total isotropic energy in the common rest frame energy band ($r=0.62$, $P=0.02$).

We define the quiescent time in the rest frame $\Delta t^{RF}$ as the separation between the $1-$s peak flux of the precursor and the subsequent emission episode. This estimate is less affected by the instrumental threshold. The quiescent time distribution in the rest frame is quite broad, from a few seconds up to $\sim 50$ s, with an average value of $18$ s. Quiescent times moderately correlate with the duration of the following emission episode ($r=0.64$, $P=0.02$), while no connection exists with the previous emission episode, as already reported by \citet{2001MNRAS.320L..25R} analyzing a larger sample of BATSE GRBs (see also \citealt{2009A&A...505..569B}). They concluded that a similar behavior is indicative of an accumulation mechanism such that the longer the waiting time, the higher the stored energy available for the next emission episode. A further hint comes from the moderate correlation we found between the ratio of the subsequent and the previous emission episode energy in the BAT energy band (usually precursor and main event, unless multiple precursors are present) and the rest frame waiting time between the two ($r=0.60$, $P=0.03$).

\subsection{The multiwavelength properties of GRBs with and without precursors in BAT6}\label{afterglow}

We compared the multiwavelength properties of GRBs with precursors (superscript ``p'') with those of GRBs without precursors in the BAT6 sample (superscript ``no-p'') in order to single out differences that may discriminate between the two groups, if any\footnote{The multiwavelength properties of GRBs in the BAT6 sample can can be retrieved in \citet{2012ApJ...749...68S,2012MNRAS.421.1697C,2013arXiv1303.4743C,2012MNRAS.425..506D,2012MNRAS.421.1265M,2012MNRAS.421.1256N}.}.

We first compared the prompt emission properties of GRBs with and without precursors in the BAT6 sample. No sharp distinction exists in the distributions of the peak energy $E_{\rm pk}$ (a KS test gives a probability $P=1$ that the two samples are drawn from the same parent population), total isotropic energy $E_{\rm \gamma,iso}$ ($P=0.41$) or peak luminosity $L_{\rm \gamma,pk}$ ($P=0.92$) (see \citealt{2012MNRAS.421.1256N} for the definition of these quantities). We conclude that it is not possible to predict the presence of a precursor from the prompt emission properties only. 

Comparing the redshift distribution of GRBs with and without precursors in the BAT6 sample with a KS test, we found a probability $P= 0.40$ that they are drawn from the same population. This suggests that the progenitors of GRBs with and without precursors may belong to the same population.

We then turn to the properties of the X--ray emission. We restrict our analysis to those GRBs with a plateau in the X--ray light curve (the ``canonical'' light curves, \citealt{2006ApJ...642..389N}) of GRBs with and without precursor in the BAT6. In particular, for those GRBs with redshift, we compared the plateau rest frame properties. The X--ray luminosity is calculated in a common rest frame $0.3-30$ keV energy band \citep{2013MNRAS.428..729M}, representative of the bolometric luminosity. The central value (median) of the luminosity at the end of the plateau is larger for the GRBs with precursors than that for the ones without precursors in the BAT6 ($\langle log[L_{\rm end}^{\rm p}]\rangle=47.7$ while $\langle log[L_{\rm end}^{\rm no-p}]\rangle=46.2$), while the dispersion is much smaller ($\sigma_{\rm p}=0.3$ while $\sigma_{\rm no-p}=1.5$). A KS test gives a probability $P=0.02$ that the two distributions are drawn from the same population. No clear differences emerge in the distributions of the temporal decay indices or the break times (a KS test gives a probability $P=0.97$ and $P=0.50$ that they are drawn from the same population, respectively).

\section{Quiescent times: a propeller phase?}\label{sect_propeller}

All the observed properties of GRBs with precursors that we reported in our analysis can be explained if the central GRB engine is a newly born magnetar. This proposal is based on the assumption that the GRB prompt emission originates from a newly born magnetar accreting material from an accretion disk, and the observed power is proportional to the mass accretion rate. Close to the surface of the magnetar, the behavior of the in--falling material is dominated by the large magnetic field of the neutron star, so that matter is channelled along the field lines onto the magnetic polar caps. We considered a pure dipolar magnetic field. Higher order multipoles have a lower influence on the in--falling matter motion due to their steeper radial dependence. The magnetic field begins to dominate the motion of matter at the magnetospheric radius $r_{\rm m}$, defined by the pressure balance between the magnetic dipole of the magnetar and the in--falling material. Accretion onto the surface of the magnetar proceeds as long as the material in the disk rotates faster than the magnetosphere (see Fig.~1, right upper panel). In the opposite case, accretion can be substantially reduced due to centrifugal forces exerted by the super-Keplerian magnetosphere: the source is said to enter the ``propeller'' phase \citep{1975A&A....39..185I,1998A&ARv...8..279C} (see Fig.~1, right lower panel).

\begin{figure*}
\includegraphics[width=\hsize,clip]{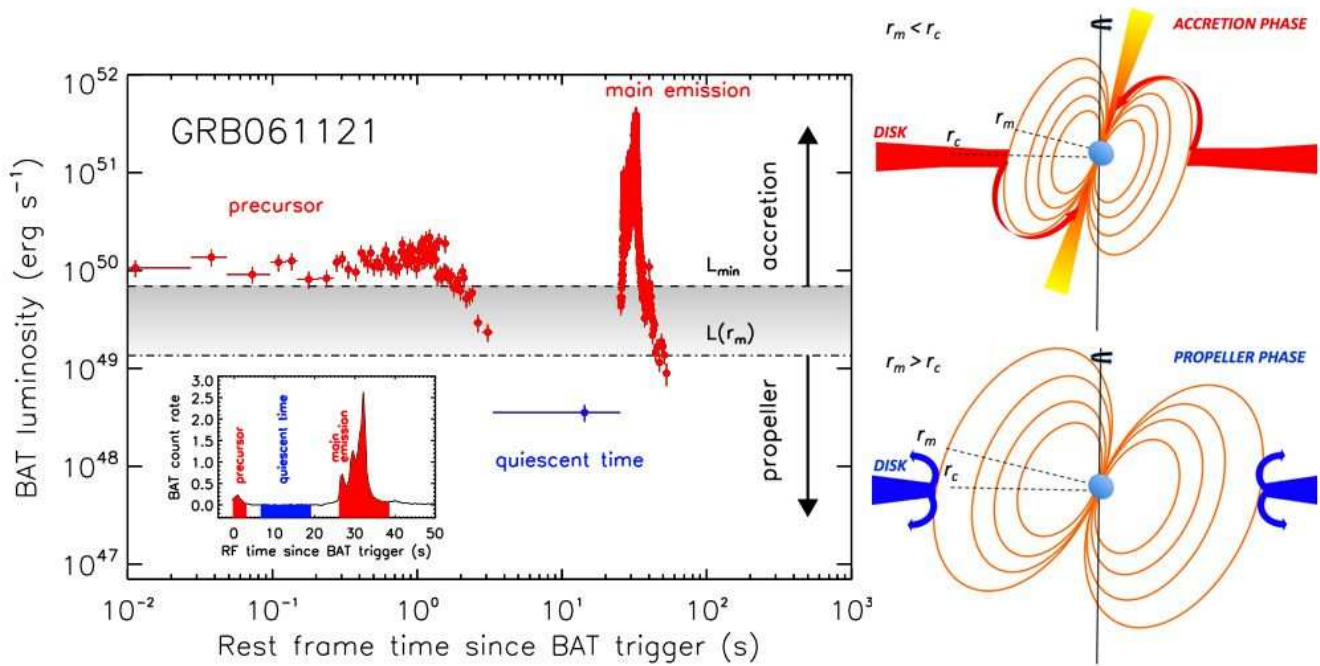}
\caption{\textit{Left panel}: $15-150$ keV luminosity of GRB 061121 binned with signal-to-noise $S/N=5$. The minimum luminosity before the onset of the propeller phase $L_{\rm min}$ (dashed line) and the maximum quiescent time luminosity $L(r_{\rm m})$ (dotted line) are compared with the precursor (red dots), quiescence (blue dots) and main event (red dots) emission.  Characteristic luminosities have been derived independently from the fitting of the late--time X--ray emission with the spin--down power of the magnetar. The values derived for the magnetic field and spin period with this procedure are: $B=6.03\times 10^{15}$ G, $P=4.40$ ms. We assumed a beaming factor $f_b=0.01$ and a radiative efficiency $\epsilon_r=0.1$. \textit{Inset}: BAT count rate light curve of GRB 061121, with the precursor (red area), the quiescence (blue area) and the main event (red area). 
\textit{Right upper panel:} accretion onto the surface of the magnetar. The magnetospheric radius $r_{\rm m}$ is smaller than the co--rotation radius $r_{\rm c}$, where the magnetosphere centrifugal drag balances gravity: in--falling matter from the accretion disk rotates faster than the magnetosphere and accretion onto the magnetar surface takes place. This phase accounts for both the precursor(s) and the main event emission. 
\textit{Right lower panel:} propeller phase. The magnetospheric radius $r_{\rm m}$ is larger than the co--rotation radius $r_{\rm c}$: centrifugal forces on the in--falling matter at $r_{\rm m}$ are too large to allow for co--rotation, the net radial force is outward and the in--fall velocity drops to zero as well as the accretion power. This phase corresponds to the quiescent times between the precursor(s) and the main emission. Since $r_{\rm m} \propto \dot{M}^{-2/7}$, when enough matter is accumulated to fulfill the condition $r_{\rm m}< r_{\rm c}$ the propeller phase ends and accretion restarts, corresponding to a new emission episode.}
\end{figure*}

In this scenario, during the collapse of the progenitor, a proto--magnetar is formed and the confining pressure and inertia of the in--falling stellar envelope act to collimate the outflow into a jet \citep{2007ApJ...669..546U}. This matter forms a disk and accretes onto the newly born magnetar, giving rise to the GRB power. If the conditions are profitable the system can enter the propeller phase: accretion is inhibited and the GRB becomes quiescent. During this phase, matter continues to pileÐup at $r_{\rm m}$  (at a few neutron star radii) until its pressure is high enough to overcome the centrifugal barrier again. Accretion onto the surface of the neutron star then restarts, giving rise to another high energy event. All the emission episodes are produced by the same mechanism and, thus, have the same observational properties.  

Every emission episode should lie above the characteristic luminosity corresponding to the onset of the propeller phase $L_{\rm min}$, providing a strong observational test for the consistency of this model. This is indeed confirmed for the GRBs with precursors in the BAT6 sample, with typical values for the magnetic field and the spin period of a newly born magnetar \citep{1992ApJ...392L...9D} (see Fig.~1, left panel where GRB 061121 is portrayed as an example, and Table~\ref{tab_magnetar}). It is possible to have multiple precursors, if the centrifugal barrier is penetrated more than once, i.e. if centrifugal forces are weaker than for single precursor emission. This is the case for, e.g., GRB 070306 (see Appendix~\ref{app_magnetar}).

During the propeller phase the luminosity does not drop to zero. A smaller luminosity $L(r_{\rm m})$ is expected resulting from the gravitational energy release of the in--falling matter up to $r_{\rm m}$, escaping through the pre-excavated funnel. This provides an upper limit to the observed quiescent time luminosity since only a fraction of it may leak out from the jet base. The \textit{Swift}/BAT was triggered by the precursor of GRB 060124 \citep{2006A&A...456..917R}, a burst that displayed a very long quiescent time ($\sim 90$ s in the cosmological rest frame), allowing the \textit{Swift} satellite to slew to the source and detect the quiescent emission with the X--Ray Telescope (XRT, \citealt{2005SSRv..120..165B}). This quiescent luminosity is almost constant and $\sim 100$ times fainter than the main event, well below the predictions for $L(r_{\rm m})$ (see Fig.~2, left panel).

An expectation for the propeller model is that the longer the quiescent time, the stronger and longer the subsequent emission episode, due to the larger amount of matter stored during the propeller phase. For the special cases of GRB 070306 and GRB 061222A with two precursors\footnote{The other GRBs with multiple precursors are GRB 060904A and GRB 080602, where the multiple precursors are not well separated in time.}, these scalings can be tested within the same burst. In particular we found that in GRB 061222A both the total energy emitted in the BAT energy band and the duration ($T_{90}$\footnote{Time interval between $5\%$ and $95\%$ of the fluence being emitted.}) are proportional to the previous quiescent time for the second precursor and the main event, that is what we expect from an accumulation mechanism at work, such as the propeller one. Precursors and main event in GRB 070306 are separated by comparable time intervals, while the energetic of the main event is a factor three larger than in the previous precursor. This might indicate that, at variance with the previous case, the mass accretion rate is not constant during the evolution of this burst. A correlation between the quiescent time and the subsequent emission episode duration was reported in the BATSE sample \citep{2001MNRAS.320L..25R}. 

Being time-symmetric, the propeller model can still work after the main event, producing ``postcursors''. The giant flares observed in the X--ray light curve of the GRB afterglows (e.g., GRB 050502B, \citealt{2005Sci...309.1833B}) could be explained in this way.

\begin{figure*}
\includegraphics[width=\hsize,clip]{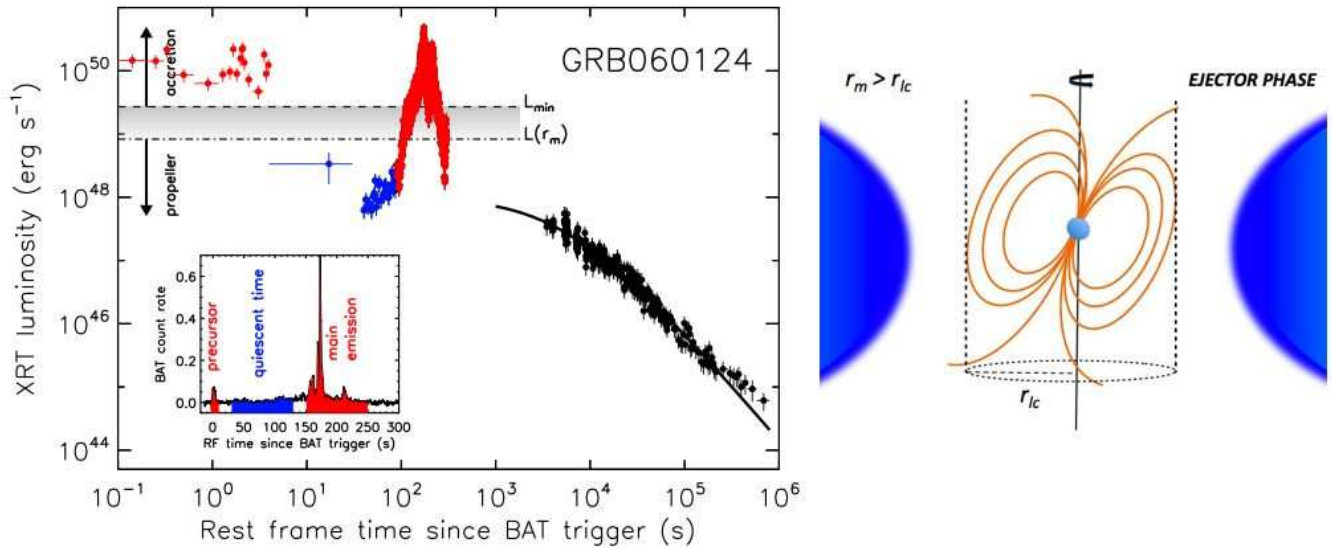}
\caption{\textit{Left panel}: $0.3-30$ keV luminosity of GRB 060124. Luminosity lines are compared with the main event emission in the X--ray band, as well as with the precursor emission rescaled in the $0.3-30$ keV energy band: the main event emission (red points) and the precursor emission (red points) are consistent with accretion onto the magnetar surface, while the quiescent time emission (blue points) is below the estimate for $L(r_{\rm m})$. The late time X--ray afterglow emission (black points) has been fitted assuming that the spin--down power emitted by the magnetar is the source of energy injected in the forward shock. This produces the shallow decay in the X-ray light curve \citep{2011A&A...526A.121D} (black solid line), giving a direct estimate of the magnetar magnetic field and spin period: $B=1.93\times 10^{15}$ G, $P=2.08$ ms. We assumed a beaming factor $f_b=0.01$ and a radiative efficiency $\epsilon_r=0.1$. \textit{Inset}: BAT count rate light curve of GRB 060124, showing the precursor (red area), the quiescence (blue area) and the main event (red area). 
\textit{Right panel}: ejector phase. As the mass inflow rate decreases further, the magnetospheric radius expands reaching the light cylinder, $r_{\rm lc}$. Beyond the light cylinder the field becomes radiative, dominating the in--falling material pressure at any distance from the central magnetar. As a consequence, the accreting matter is pushed outwards preventing from other accretion episodes. This sets the termination of the accretion phase and the start of the dipole rotational losses phase.}
\end{figure*}

The accretion process ends when the mass inflow rate decreases enough for the magnetospheric radius to reach the light cylinder (i.e. the radius at which the field lines co--rotate with the neutron star at the speed of light, see Fig~2, right panel). Beyond this radius the field becomes radiative and expels much of the in--falling matter. After the end of the prompt emission, the GRB afterglow can also be influenced by the magnetar, being re--energized by its spin--down power \citep{2011A&A...526A.121D}. We applied the solution of \citet{2011A&A...526A.121D} to our sample (details are given in Appendix~\ref{app_magnetar}) to have a direct and independent estimate of  the magnetic field and spin period of the magnetar ($B\sim (2-20)\times 10^{15}$ G, $P\sim 2-10$ ms) and to calculate the characteristic luminosities of the propeller phase for different GRBs. In all cases we find that the precursor and main GRB event occur in the accretion phase, whereas the luminosity between them falls in the propeller phase. The estimate of the spin period from late--time emission may be biased by spin--up and down occurring during the accretion and propeller phases. However, it has been demonstrated that the spin period does not change appreciably assuming reliable mass--accretion rate during supernova fallback onto a magnetar \citep{2011ApJ...736..108P}.

If the initial mass inflow rate is not high enough to shut off the spin--down power, a magnetar powered GRB can occur. In this case we might expect a smoother and longer event, such as GRB 060218 \citep{2006Natur.442.1008C,2006Natur.442.1014S}. At the other extreme, if during the accretion phase the magnetar accretes enough matter, then it collapses to a black hole \citep{2011ApJ...736..108P}. A clear signature of this transition to a black hole is the absence of the spin--down power emitted by the magnetar able to re--energize the late--time X--ray emission and thus no plateau in the X-ray afterglow should emerge. An example of this phenomenon in the BAT6 sample is provided by GRB 061007 \citep{2007MNRAS.380.1041S}, which shows precursor activity and a single power--law decaying X--ray afterglow. When the prompt and precursor energies are converted to accreted mass, GRB 061007 is found to be the object that has accreted the largest amount of mass among the GRBs with precursors in the BAT6 sample (see  Appendix~\ref{app_magnetar}).

\section{The rate of magnetars and GRBs}\label{sect_rate}

The local rate of core--collapse Supernovae (SNe) is $\sim 0.01$ yr$^{-1}$ \citep{2011LRR....14....1F}. Core--collapse SNe can be mainly divided into type II (usually associated to supergiant progenitors) and type Ibc (usually associated to Wolf-Rayet progenitors). Sn Ibc are $\sim 30\%$ of core--collapse SNe \citep{2011MNRAS.412.1473L}, leading to a local rate for the SNe Ibc $\approx 3\times 10^{-3}$ yr$^{-1}$. Magnetars constitute about $10\%$ of the Galactic neutron star birth-rate \citep{2005ApJ...620L..95G,2007MNRAS.381...52G,2008ApJ...680..639M,1994Natur.368..125K}, and it is expected that a similar fraction of SNe Ibc give rise to a magnetar \citep{2005ApJ...620L..95G,2006Natur.442.1014S}, i.e. $\sim 10\%$ of SNe Ibc host a magnetar. 

Long GRBs in the local Universe have been unambiguously associated to SNe Ibc. The SNe Ibc associated to GRBs have broad emission lines and are extremely energetic (called Hypernovae, \citealt{1998Natur.395..672I}). At the same time not all broad-lines SNe Ibc are related to GRBs (e.g. SN 2002ap, \citealt{2002ApJ...572L..61M}), likely indicating that there is the need for an additional ingredient. SNe Ibc were initially associated to low--luminosity GRBs (with peak luminosity $L< 10^{49}$ ergs s$^{-1}$, such as GRB 980425/SN 1998bw, \citealt{1999A&AS..138..463P}, or GRB 060218/SN 2006aj, \citealt{2006Natur.442.1008C,2006Natur.442.1014S}), but later on they were discovered also in normal (i.e. the ones considered in our analysis) GRBs (such as GRB 030329/SN 2003dh, \citealt{2003ApJ...591L..17S,2003Natur.423..847H}).

The rate of GRBs has been constrained by independent and consistent estimates. The rate of SNe Ibc with an associated low--luminosity GRB is between $\sim 1\%-9\%$, while those associated to a normal GRB are about $0.3\%-3\%$, depending on the assumption on the beaming factor \citep{2006Natur.442.1014S,2007ApJ...657L..73G,2013MNRAS.428.1410G}. Estimates from the radio surveys give that the incidence of low--luminosity events is $< 3\%$ in SNe Ibc \citep{2003ApJ...599..408B}, while the limit on the fraction of SNe Ibc actually associated with a GRB, including off--axis events, is $< 10\%$ \citep{2006ApJ...638..930S}. 

It is clear from the above estimates that, overall, magnetars related to SNe Ibc are consistent with the total number of observed GRBs, accounting for both low--luminosity and normal GRBs.

\section{Conclusions}\label{conclusions}

The main observational properties of the GRBs with precursors can be qualitatively explained if the central GRB engine is a newly born magnetar and the precursor and the prompt emission arise from the accretion of matter onto its surface. In this scenario the accretion process can be halted by the centrifugal drag exerted by the rotating magnetosphere onto the in--falling matter, allowing for multiple precursors and very long quiescent times.

Using the complete BAT6 sample to estimate the fraction of long GRBs with precursors in an unbiased way, we end up with $\sim17\%$ of GRBs showing precursor activity. The fraction of GRBs powered by magnetar is even higher if we interpret the X--ray plateau in the GRB afterglows as due to continuous  energy injection from the magnetar spin--down power, resulting in  at least $\sim 80\%$. The remaining GRBs can be powered by a newly born black hole or by a magnetar collapsed to a black hole (GRB 061007 alike). Indeed, the collapse of a massive star accommodates both the direct collapse to a black hole and the formation of a proto--magnetar in those cases where fast--rotating cores produce a magneto-rotational explosion. Despite the uncertainties, recent simulations seem to indicate that proto--magnetars are more easily produced than black holes by current stellar--evolutionary models \citep{2012ApJ...754...76D}. Within current uncertainties, the rate of magnetar--powered SN Ibc is comparable to the rate of (low-luminosity and normal) GRBs, indicating that a large fraction of the long GRBs are powered by a magnetar.

The propeller mechanism as an explanation for the quiescent time could also be extended to short GRB precursors \citep{2010ApJ...723.1711T}. In this case the central magnetar would originate from a binary white dwarf merger, a white dwarf accretion induced collapse, or perhaps, the merger of a double neutron star binary \citep{2006MNRAS.368L...1L,2008MNRAS.385.1455M}.

\acknowledgments

The authors thank the referee for his/her useful comments. The authors acknowledge support from ASI grant I/004/11/0 and PRIN-MIUR 2009 grants. DB is funded through ARC grant DP110102034. MGB and S. Campana want to thank L. Stella and S. Dall'Osso for conversations, P. Romano and R. Margutti for having supplied data of GRB 060124, A. Hawken for language editing.

\bibliographystyle{aa}

\bibliography{precursor}

\appendix

\section{Late--time X--ray emission and the estimate of the parameters of the magnetar}\label{app_magnetar}

The observation of a flattening in the X--ray light curve (plateau) in a large fraction of GRBs ($46 \%$ in the BAT6 sample) can be explained as an injection of energy into the forward shock (the GRB afterglow, \citealt{2006ApJ...642..354Z}). This fraction is even larger ($80 \%$ in the BAT6 sample) if we include also those GRBs displaying a shallow decay phase without the initial steep decay \citep{2012A&A...539A...3B,2012MNRAS.425..506D}. A natural source for this energy is the power emitted by a spinning--down newly born magnetar \citep{1998A&A...333L..87D,2001ApJ...552L..35Z,2009ApJ...702.1171C,2011A&A...526A.121D}. This proposal has been successfully tested both for long \citep{2010MNRAS.402..705L,2011A&A...526A.121D,2012A&A...539A...3B} and short\citep{2013arXiv1301.0629R} GRBs. In particular, the plateau luminosity and its temporal duration are directly related to the spin--down luminosity and timescale, and, thus, to the magnetic field ($B$) and the spin period ($P$) of the magnetar. The analysis of the plateau phase in the X--ray light curves provides a direct estimate of these parameters.

We refer to the model proposed by \citet{2011A&A...526A.121D}, that calculated analytically the contribution to the forward shock of the power emitted by a millisecond spinning, ultramagnetized neutron star at time $t$ as:
\begin{equation}
\frac{dE(t)}{dt}=L_{\rm sd}(t)-k' \frac{E(t)}{t}=\frac{L_i}{(1+at)^2}-k' \frac{E(t)}{t}\, ,
\end{equation}
where $L_i=IB^2R^6/(6Ic^3P^{4})\propto B^2/P^4$ is the initial spindown luminosity ($I$ is the moment of inertia, $R$ the radius of the magnetar\footnote{Here and in what follows we assume for the mass of the magnetar $M=1.4 M_{\odot}$ and for the radius $R=10^6$ cm.}, $c$ the speed of light), $a=1/t_{b2}=2B^2R^6/(6Ic^3P^{2})\propto B^2/P^2$ is the inverse of the spindown timescale $t_{b2}$, $E$ the forward shock energy, and $k'$ is a parameter that accounts for our ignorance about the microphysical parameters and on the density profile of the ambient medium (in general $0<k'<1$). A solution of this equation is:
\begin{equation}
E(t)=\frac{L_i}{t^{k'}}\int^t_{t_\circ} \frac{t^{k'}}{(1+at)^2}+E_\circ \left( \frac{t_\circ}{t} \right)^{k'}\, ,
\end{equation}
where $t_\circ$ is any time chosen as initial condition and $E_\circ$ the initial energy. The solution of the above integral can be expressed in terms of the real valued hypergeometric function $_{2}F_1(a,b,c;(1+at)^{-1})$. The total bolometric luminosity is, then:
\begin{equation}
L(t)=E(t)/t\, .
\label{lum}
\end{equation}

\begin{figure}
\centering
\includegraphics[width= \hsize,clip]{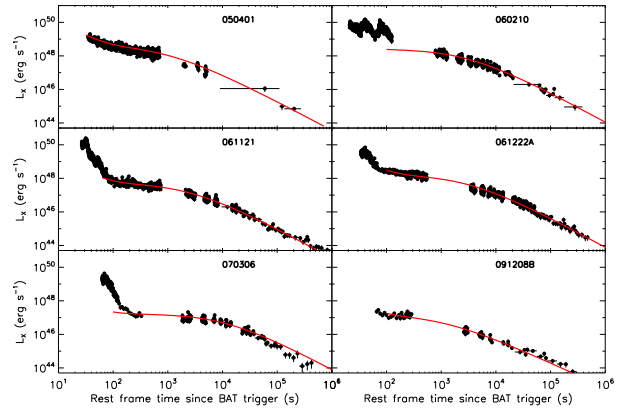}
\caption{$0.3-30$ keV luminosity of the GRBs in the BAT6 sample with precursors (``p'') and displaying a plateau phase in the X--ray light curve, fitted with the model in Eq.~\ref{lum} (red line) for the best-fit values reported in Table~\ref{tab_magnetar}.}
\label{fit_magnetar}
\end{figure}

We selected those GRBs in the BAT6 sample with redshift and with a well--sampled plateau in the X--ray light curve ($16$ GRBs), having or not a precursor in the prompt emission, and we assumed their $0.3-30$ keV common rest frame luminosity \citep{2013MNRAS.428..729M} as a proxy of the total bolometric luminosity. In order to account for the possible collimation of the outflow $\theta_j$ and of the radiative efficiency $\epsilon_r$, we considered the corrected luminosity $L_{\rm X,j}= (f_b/\epsilon_r)\,L_{\rm X,iso}$, with $\epsilon_r=0.1$ and $f_b=(1-\cos \theta_j)= 0.01$, that corresponds to $\theta_j\simeq 8^\circ$. We fitted these data with Eq.~\ref{lum}, using as free parameters $B$, $P$ and $E_\circ$. We fixed $t_\circ$ as the (rest--frame) starting time of the plateau phase, and $k'$ from the decay index of the post--plateau light curve (the solution in Eq.~\ref{lum} has an asymptotic behavior $\propto t^{-k'-1}$, for detail see \citealt{2011A&A...526A.121D}).  Tables~\ref{tab_magnetar} and \ref{tab_magnetar2} summarize the best--fit parameters for the GRBs, grouped as GRBs with precursors and without precursors, respectively, while Figure~\ref{fit_magnetar} shows the results of the fit for the sample of GRBs with precursors only.

\begin{deluxetable*}{cc|ccccc|ccc}
\tablecaption{Best--fitting values of the plateau phase for the GRBs with precursors (superscript ``p'') in the BAT6 sample. \label{tab_magnetar}}
\tablewidth{0pt}
\tablehead{\colhead{Name} & \colhead{$z$} & \colhead{$B^{\rm p}$} & \colhead{$P^{\rm p}$} & \colhead{$k'^{\rm p}$} & \colhead{$t_\circ^{\rm p}$} & \colhead{$E_\circ^{\rm p}$} & \colhead{$L_{\rm min}^{\rm p}$} & \colhead{$L^{\rm p}(r_{\rm m})$} & \colhead{$M_{\rm acc}^{\rm p}$}\\
\colhead{} & \colhead{} & \colhead{($10^{15}$ G)} & \colhead{(ms)} & \colhead{} & \colhead{(s)} & \colhead{($10^{50}$ erg)} & \colhead{($10^{50}$ erg s$^{-1}$)} & \colhead{($10^{50}$ erg s$^{-1}$)} & \colhead{($M_\odot$)}}
\startdata
050318\tablenotemark{a} & $1.44$ & $4.00$ & $3.06$ & $-$& $-$& $-$& $4.7$& $1.1$ & $0.004$\\
050401 & $2.90$ & $5.67\pm0.27$ & $2.61\pm0.04$ & $0.8$ & $31.6$ & $5.57\pm0.18$ & $13.7$ & $3.6$ & $0.101$\\
060210 & $3.91$ & $2.34\pm0.07$ & $1.83\pm0.02$ & $0.9$ & $15.8$ & $1.00$\tablenotemark{b} & $5.4$ & $1.8$ & $0.088$\\
061007\tablenotemark{a} & $1.26$ & $4.00$ & $3.06$ & $-$& $-$& $-$& $4.7$& $1.1$ & $0.252$\\
061121 & $1.31$ & $6.03\pm0.12$ & $4.40\pm0.03$ & $0.9$ & $63.1$ & $0.70\pm0.04$ & $4.6$ & $0.9$ & $0.068$\\
061222A & $2.09$ & $2.79\pm0.04$ & $2.25\pm0.01$ & $0.9$ & $31.7$ & $5.06\pm0.37$ & $4.7$ & $1.4$ & $0.056$\\
070306 & $1.50$ & $2.33\pm0.10$ & $3.60\pm0.05$ & $0.9$ & $15.8$ & $0.32$\tablenotemark{b} & $1.1$ & $0.2$ & $>0.022$\\
091208B & $1.06$ & $18.6\pm1.1$ & $9.70\pm0.21$ & $0.5$ & $19.9$ & $0.36\pm0.07$ & $6.9$ & $0.8$ & $0.005$\\
\enddata
\tablecomments{Name and redshift ($z$); magnetic field ($B$) and spin period ($P$) of the magnetar, parameter that accounts for our ignorance about the microphysical parameters and on the density profile of the ambient medium ($k'$), shallow decay onset rest-frame time ($t_\circ$), initial energy of the forward shock ($E_\circ$); minimum bolometric accretion luminosity necessary to penetrate the centrifugal barrier ($L_{\rm min}$) and bolometric accretion luminosity just after the onset of the centrifugal barrier ($L(r_{\rm m})$), calculated with the best-fitting parameters $B$ and $P$, mass accreted during the prompt emission ($M_{\rm acc}$). }
\tablenotetext{a}{These GRBs do not allow to perform the fit of the late-time X--ray emission with the model in Eq.~\ref{lum}. We therefore assumed as $B$ and $P$ the median values of their distributions, and calculated the luminosities accordingly.}
\tablenotetext{b}{Not a free parameter.}
\end{deluxetable*}

\begin{deluxetable*}{cc|ccccc}
\tablecaption{Best--fitting values of the plateau phase for the GRBs without precursors (superscript ``no-p'') in the BAT6 sample. \label{tab_magnetar2}}
\tablewidth{0pt}
\tablehead{\colhead{Name} & \colhead{$z$} & \colhead{$B^{\rm no-p}$} & \colhead{$P^{\rm no-p}$} & \colhead{$k'^{\rm no-p}$} & \colhead{$t_\circ^{\rm no-p}$} & \colhead{$E_\circ^{\rm no-p}$}\\
\colhead{} & \colhead{} & \colhead{($10^{15}$ G)} & \colhead{(ms)} & \colhead{} & \colhead{(s)} & \colhead{($10^{50}$ erg)}}
\startdata
060306 & $1.55$ & $9.40\pm0.90$ & $7.30\pm0.20$ & $0.9$ & $31.6$ & $0.63\pm0.09$\\
060814 & $1.92$ & $2.87\pm0.10$ & $3.79\pm0.04$ & $0.9$ & $316.2$ & $2.22\pm0.27$\\
061021 & $0.35$ & $16.97\pm0.48$ & $32.21\pm2.90$ & $0.5$ & $316.2$ & $0.02\pm0.01$\\
080430 & $0.77$ & $5.06\pm0.21$ & $13.42\pm0.21$ & $0.9$ & $3.2$ & $0.04\pm0.01$\\
080607 & $3.04$ & $10.59\pm0.57$ & $3.61\pm0.10$ & $0.9$ & $100.0$ & $5.71\pm0.22$\\
081007 & $0.53$ & $12.72\pm0.77$ & $23.58\pm0.50$ & $0.9$ & $100.0$ & $0.06\pm0.01$\\
081203A & $2.10$ & $19.16\pm0.69$ & $5.21\pm0.10$ & $0.9$ & $31.6$ & $5.75\pm0.30$\\
081221 & $2.26$ & $7.68\pm0.25$ & $2.20\pm0.01$ & $0.9$ & $19.9$ & $30.13\pm1.73$\\
091020 & $1.71$ & $12.54\pm0.36$ & $6.64\pm0.07$ & $0.7$ & $31.7$ & $2.18\pm0.05$\\
100621A & $0.54$ & $2.95\pm0.11$ & $6.92\pm0.01$ & $0.7$ & $100.0$ & $0.74\pm0.03$\\
\enddata
\tablecomments{Name and redshift ($z$); magnetic field ($B$) and spin period ($P$) of the magnetar, parameter that accounts for our ignorance about the microphysical parameters and on the density profile of the ambient medium ($k'$), shallow decay onset rest-frame time ($t_\circ$), initial energy of the forward shock ($E_\circ$).}
\end{deluxetable*}

\subsection{The properties of the magnetar in GRBs with and without precursors.}

\begin{figure}
\centering
\includegraphics[width= \hsize,clip]{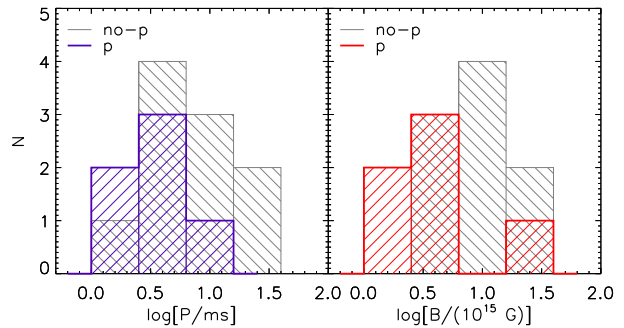}
\caption{\emph{Left panel:} spin period $P$ distribution for the GRBs in the BAT6 sample displaying a plateau phase in the X--ray light curve and with precursors (blue; ``p'', see Table~\ref{tab_magnetar}) or without precursors (gray; ``no--p'', see Table~\ref{tab_magnetar2}). \emph{Right panel:} magnetic field $B$ distribution for the GRBs in the BAT6 sample displaying a plateau phase in the X--ray light curve and with precursors (red; ``p'', see Table~\ref{tab_magnetar}) or without precursors (gray; ``no--p'', see Table~\ref{tab_magnetar2}).}
\label{distrib}
\end{figure}

Since the condition for the onset of the propeller phase depends on $B$ and $P$, we searched for a pattern that allows us to discriminate between GRBs with (superscript ``p'') and without (superscript ``no-p'') precursors. For this reason we compared the distributions of $B$ and $P$ for GRBs with and without precursors in the BAT6 sample (see Tables~\ref{tab_magnetar} and \ref{tab_magnetar2}, respectively). We find that both the magnetic field and the spin period distributions are centered around lower values for GRBs with precursors than for GRBs without precursors ($\langle log[B^{\rm p}/10^{15}$G$]\rangle=0.60$ while $\langle log[B^{\rm no-p}/10^{15}$G$]\rangle=1.00$; $\langle log[P^{\rm p}/$ms$]\rangle=0.48$ while $\langle log[P^{\rm no-p}/$ms$]\rangle=0.75$, see Figure~\ref{distrib}). The spin period distribution is also less scattered around its central value than in the other case ($\sigma^{\rm p}_{\rm P}=0.26$ while $\sigma^{\rm no-p}_{\rm P}=0.38$). A KS test gives a probability $P=0.12$ that the spin period of GRBs with and without precursors are drawn from the same population, while $P=0.30$ for the magnetic field distributions.

The luminosity of the shallow decay phase is related to the spin--down luminosity, being $L_i\propto B^2/P^4$. A lower value of the spin period and of the magnetic field for the GRBs with precursors would result in a higher luminosity during the shallow decay phase since the luminosity depends strongly on $P$. Similarly, the narrower distribution of $P$ would imply a narrower distribution of $L_i$, that is indeed what we found in Sect.~\ref{afterglow}  \citep[see also][]{2013A&A...552L...5P}. The magnetospheric radius depends on the magnetic field and on the mass accretion rate (given the mass and radius of the magnetar), $r_m\propto \dot{M}^{-2/7}B^{4/7}$, while the corotation radius depends only on the spin period, being $r_c\propto P^{2/3}$. Thus, among the GRBs powered by a magnetar, GRBs with precursors are characterised by specific values of the magnetic field and spin period that favor the trigger of the propeller regime, responsible for the observed quiescent times.

\section{The estimate of the characteristic luminosities of the propeller regime.}

We used the best--fit values of $B$ and $P$ in the case of GRBs with precursor (see Table~\ref{tab_magnetar}) to estimate in the case of spherical accretion the bolometric accretion power corresponding to the onset of the propeller phase:
\begin{equation}
L_{\rm min}=4\times 10^{50} B_{15}^2\, P_{-3}^{-7/3}\, {\rm erg\,s^{-1}}\,,
\label{Lmin}
\end{equation}
and the bolometric power during the quiescent time:
\begin{equation}
L(r_{\rm m})=2\times 10^{50} B_{15}^2\, P_{-3}^{-3}\, {\rm erg\,s^{-1}}\,,
\label{Lcor}
\end{equation}
where $B=10^{15} B_{15}$ G and $P=10^{-3} P_{-3}$ s (for details see \citealt{1998A&ARv...8..279C}, their Eq.~4 and 6; see also \citealt{2008ApJ...683.1031B} and \citealt{2011ApJ...736..108P}). Equation~\ref{Lcor} sets an upper limit to the quiescent time luminosity, since only a fraction of it will actually escape from the jet base. The values are displayed in Table~\ref{tab_magnetar}. In the case of GRB 050318 and GRB 061007, since it was not possible to estimate $B$ and $P$ from the late X--ray emission (GRB 050318 has an incomplete light curve, while GRB 061007 has a simple power-law decay), we assumed for $B$ and $P$ the median values obtained for GRBs with precursors $B=4.00\times 10^{15}$ G and $P=3.09$ ms. 

For a direct comparison with the prompt emission as observed by BAT (see Sect.~\ref{sect_bat}), we computed the corresponding observed flux from $L_{\rm min}$: $f_{\rm min}=\frac{K L_{\rm min}}{(f_b/\epsilon_r) 4 \pi D_L^2(z)}$, where $D_L(z)$ is the luminosity distance at redshift $z$ and $K$ is a K--correction to account for the limited energy band observed by BAT\footnote{for those GRBs with measured $E_{pk}$ we calculated the fraction of energy emitted in the observed $15-150$ keV with respect to the bolometric rest-frame $1-10^4$ keV, while for the other cases we assumed this fraction to be $0.5$.}. The results are displayed in Table~\ref{tab_bat}: the main event peak flux is always well above $f_{\rm min}$, as well as most of the precursor(s) peak flux(es). In the worst cases, the precursor peak flux is comparable to $f_{\rm min}$. Indeed, the smallest $L_{\rm min}$ is found for GRB 070306, that has two well separated precursors. In this case, weaker centrifugal forces allow for multiple onsets of accretion episodes. 

We estimated also the total amount of mass accreted during the prompt emission (precursor(s) plus main emission) from the total bolometric prompt emission energy, corrected for the same collimation factor: $E_{\rm \gamma,j}=(GM_{\rm NS}/R_{\rm NS})\,M_{\rm acc}$ (see Table~\ref{tab_magnetar}; for $E_{\rm \gamma,iso}$ we refer to Table~1 in \citealt{2012MNRAS.421.1256N}). The largest amount of mass accreted corresponds to GRB 061007, that indeed has one among the largest $E_{\rm \gamma,iso}$ in the BAT6 sample \citep{2012MNRAS.421.1256N}. 

The values we derived depend on our assumption for the ratio $(f_b/\epsilon_r)$. However, a simple scaling exists that allows to derive all the parameters for different collimation angles and efficiencies, up to the isotropic case. In fact, the isotropic spindown luminosity is $L_{i,j}=(f_b/\epsilon_r)\,L_{\rm i,iso}$, which implies $B_j^2/P_j^4=(f_b/\epsilon_r) B_{\rm iso}^2/P_{\rm iso}^4$. At the same time, the spindown timescale is not influenced by collimation, $t_{b2,{\rm iso}}=t_{b2,j}$ and therefore $P_j^2/B_j^2=P_{\rm iso}^2/B_{\rm iso}^2$. Globally, we find that $P_{\rm iso}=\sqrt{(f_b/\epsilon_r)}\,P_j$ and $B_{\rm iso}=\sqrt{(f_b/\epsilon_r)}\,B_j$. Consequently, the dependence of the minimum luminosity on $(f_b/\epsilon_r)$ is $L_{\rm min}\propto B_j^2/P_j^{7/3}=(f_b/\epsilon_r)^{1/6} B_{\rm iso}^2/P_{\rm iso}^{7/3}$.

\end{document}